\documentclass{pnastwo}
\usepackage{graphicx}
\usepackage{float}

\usepackage{amssymb,amsfonts,amsmath}

\contributor{Submitted to Proceedings
of the National Academy of Sciences of the United States of America}
\url{www.pnas.org/cgi/doi/10.1073/pnas.0709640104}
\copyrightyear{2013}
\issuedate{Issue Date}
\volume{Volume}
\issuenumber{Issue Number}

\begin{document}



\title{Microwave conductance in random waveguides in the crossover to Anderson localization and single parameter scaling}




\author{Zhou Shi\affil{1}{Department of Physics, Queens College of the City University of New York, Flushing, NY 11367, USA}\affil{2}{Graduate Center of the City University of New York, New York, NY 10016, USA}, Jing Wang\affil{1}{}\affil{2}{}, \and Azriel Z. Genack\affil{1}{}\affil{2}{}}

\contributor{Submitted to Proceedings of the National Academy of Sciences
of the United States of America}

\maketitle

\begin{article}

\begin{abstract}
The nature of transport of electrons and classical waves in disordered systems depends upon the proximity to the Anderson localization transition between freely diffusing and localized waves. The suppression of average transport and the enhancement of relative fluctuations in conductance in one-dimensional samples with lengths greatly exceeding the localization length, $L\gg \xi$, are related in the single parameter scaling (SPS) theory of localization. However, the difficulty of producing an ensemble of statistically equivalent samples in which the electron wavefunction is temporally coherent has so-far precluded the experimental demonstration of SPS. Here we demonstrate SPS in random multichannel systems for the transmittance $T$ of microwave radiation, which is the analogue of the dimensionless conductance. We show that for $L\sim4\xi$ a single eigenvalue of the transmission matrix (TM) dominates transmission and the distribution of the $\ln T$ is Gaussian with a variance equal to the average of $-\ln T$, as conjectured by SPS. For samples in the crossover to localization, $L\sim\xi$, we find a one-sided distribution for $\ln T$. This anomalous distribution is explained in terms of a charge model for the eigenvalues of the transmission matrix $\tau$ in which the Coulomb interaction between charges mimics the repulsion between the eigenvalues of transmission matrix. We show in the localization limit that the joint distribution of $T$ and the effective number of transmission eigenvalues determines the probability distributions of intensity and total transmission for a single incident channel. 
\end{abstract}







\dropcap{T}he suppression of transport of coherent quantum or classical waves in disordered media known as Anderson localization \cite{1,2,3} has been observed for electrons in solids \cite{4}, atoms in laser speckle patterns \cite{5}, and electromagnetic and acoustic waves in random dielectric and metallic structures \cite{6c,6a,6b,6}. The variance of relative fluctuations of conductance or of transmission of classical waves increases as $\langle T \rangle$ falls \cite{37,38,7a,6a,7}, where $\langle\dots\rangle$ indicates averaging over an ensemble of samples. Large variations in conductance relative to its average value occur for localized waves since the wave can be on- or off-resonance with modes of the medium with centers of localization that are at different positions within the sample \cite{8,9,9a}. In addition, modes of the medium may occasionally overlap to enhance coupling through the sample \cite{10}. Such fluctuations make it impossible to predict transport in an individual disordered sample. However, fluctuations of the field within disordered samples may be exploited to sharpen focusing in random systems \cite{11,12} and lower the threshold for lasing \cite{13,14}. A full account of transport in a random system would begin with the measurement of the probability distributions of conductance in ensembles of random samples with different physical dimensions in the crossover to Anderson localization. 

The importance of fluctuations in electronics was first recognized in calculations of conductance mediated by localized states \cite{3}. The first observations of the impact of mesoscopic fluctuations, however, were of universal conductance fluctuations in diffusive samples \cite{15,16,17,18} and in transmission of classical waves, such as light, microwave radiation and ultrasound \cite{6a,6b,18a}. Electronics and optics are directly connected through the Landauer relation in 1D \cite{19,20} and multichannel \cite{21} systems. This relation expresses the equivalence between the dimensionless conductance, which is the conductance $G$ in units of the quantum of conductance, and the transmittance, which is the sum of flux transmission coefficients over all incident and outgoing channels, ${\textsl g}=\langle G/(e^2/h)\rangle=\langle T\rangle=\langle \sum_{a,b=1}^N |t_{ba}|^2\rangle$. Here, the elements of the TM, $t_{ba}$, are the field transmission coefficients which relate the field in outgoing channels {\it b} to the field in incident channels {\it a}, $E_b=\sum_{a=1}^N t_{ba}E_a$ \cite{22,23}. {\it N} is the number of freely propagating channels supported by the sample leads or by the empty waveguide at a given voltage or frequency. Squaring the amplitude of the transmission coefficient $t_{ba}$ gives the intensity $T_{ba}$. Total transmission $T_a$ for an incoming channel {\it a} is equal to the sum of $T_{ba}$ over all output channels {\it b}. The localization threshold lies at ${\textsl g}=1$ \cite{2}. In 1D, $T$ is the transmission, while in multichannel systems with constant cross section, $T$ is the transmittance.

Anderson {\it et al.} \cite{3} considered the statistics of the dimensionless conductance in 1D samples with length greatly exceeding the localization length, $L/\xi \gg 1$. They hypothesized that the probability distribution of the logarithm of the dimensionless conductance in 1D is a Gaussian function with $\rm{var}(\ln {\textsl g}) approaching -\langle \ln {\textsl g}\rangle={\it L}/\xi$ for samples much longer than the localization length, $L/\xi \gg 1$, so the distribution depends upon a single parameter \cite{3}. Simulations \cite{24,25,26,27} and random matrix theory calculations \cite{28,29} find that the distribution of the conductance in diffusive multichannel systems, $L/\xi \ll 1$, is Gaussian and that the distribution of $\ln T$ becomes a one-sided log-normal function as {\textsl g} falls below unity. Measurements of the distribution of conductance have been made in ohmic samples by varying the applied voltage or magnetic field \cite{15,30}, but the corresponding measurements for localized waves have not been carried out for quantum or classical waves. 

The analogy between quantum and classical waves seen in the Landauer relation can also be expressed in terms of the eigenvalues $\tau_n$ of the matrix product $tt^\dagger$, $T=\sum_{n=1}^N \tau_n$. The measurement of the TM in random ensembles makes it possible to explore the statistics of conductance and of intensity and total transmission in a single system \cite{31}. Optical measurements of the TM have been exploited recently to focus light transmitted through strongly scattering media \cite{12}. But the dimensions of the matrix were too small relative to {\textsl g} for the impact of mesoscopic correlation to be manifest on the transmission eigenvalues and for transmission of the most transmissive channel to be substantially larger than the average transmission. 

\section{Results and Discussion}
\begin{figure}[htc]
\centering
\includegraphics[width=3.3in]{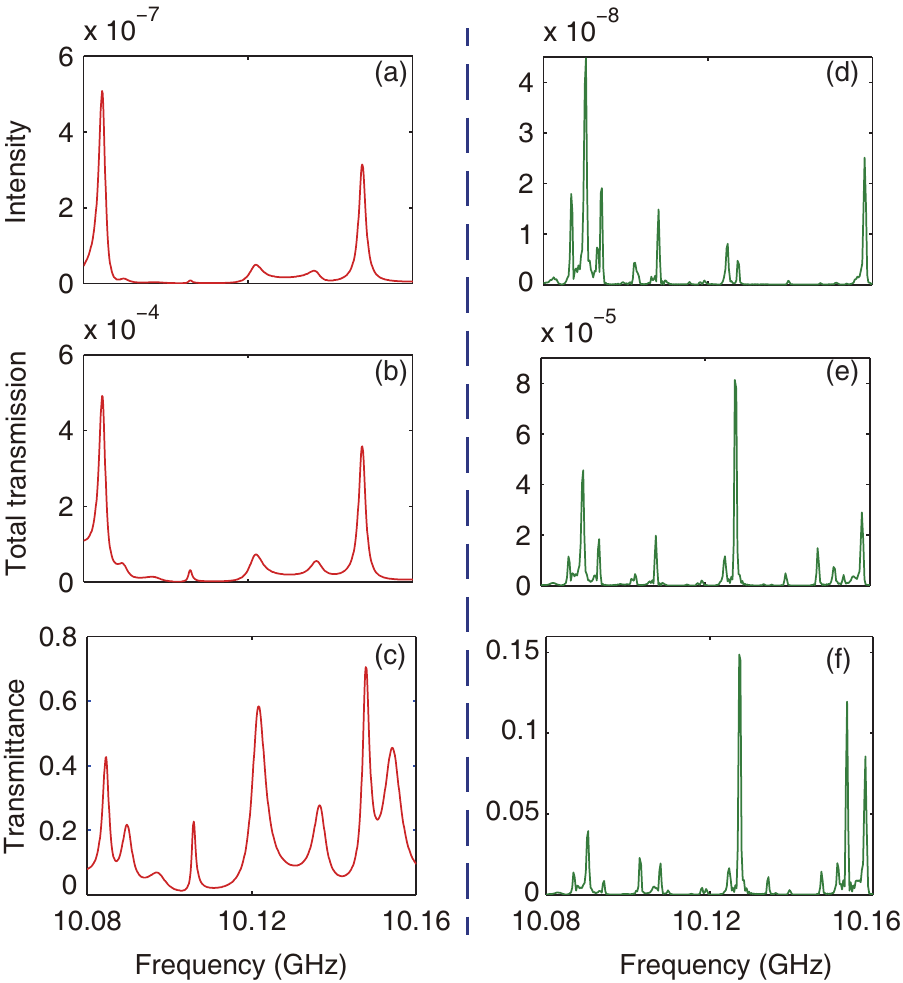}
\caption{Spectra of intensity $T_{ba}$, total transmission $T_a=\sum_{b=1}^N T_{ba}$ and transmittance $T=\sum_{a=1}^N T_a$ for microwave radiation propagating through random waveguides. Typical spectra for sample lengths $L$= 40 (left column) and 102 cm (right column) are presented. }\label{Fig1}
\end{figure}

Here we present measurements of intensity, total transmission and transmittance for localized waves in single random realizations of the sample for two different lengths in Fig. 1. The statistics for each of these transmission variables and the relationship between these statistics will be explored below. The peaks in intensity, total transmission and transmittance are seen to be more distinct in the longer and more strongly localized sample. Because the modes of the medium are well differentiated in the longer sample, sharp peaks remain even when the intensity is summed over all pairs of incident and outgoing channels. The measurement and analysis of the TM as well as the method used to remove the impact of absorption on the transmission and its statistics \cite{6a}, which has been applied to all measurements, are discussed in the {\it Materials and Methods} section.

\subsection{Statistics of conductance}
The probability distributions of $\ln T$, $P(\ln T)$, for microwave radiation transmitted through random ensembles of dielectric samples with \textsl{g}= 0.37 and 0.045 are shown in Fig. 2a. $P(\ln T)$ is nearly Gaussian for \textsl{g}= 0.045. For \textsl{g}= 0.37, the low-transmission side of $P(\ln T)$ is well fit by a Gaussian distribution, while the high transmission side falls sharply above the peak at $\ln T=-0.5$. Above $T=1.1$, $P(T)$ is seen in Fig. 2b to fall exponentially, in accord with predictions of Ref. 37. 
\begin{figure}[htc]
\centering
\includegraphics[width=3.2in]{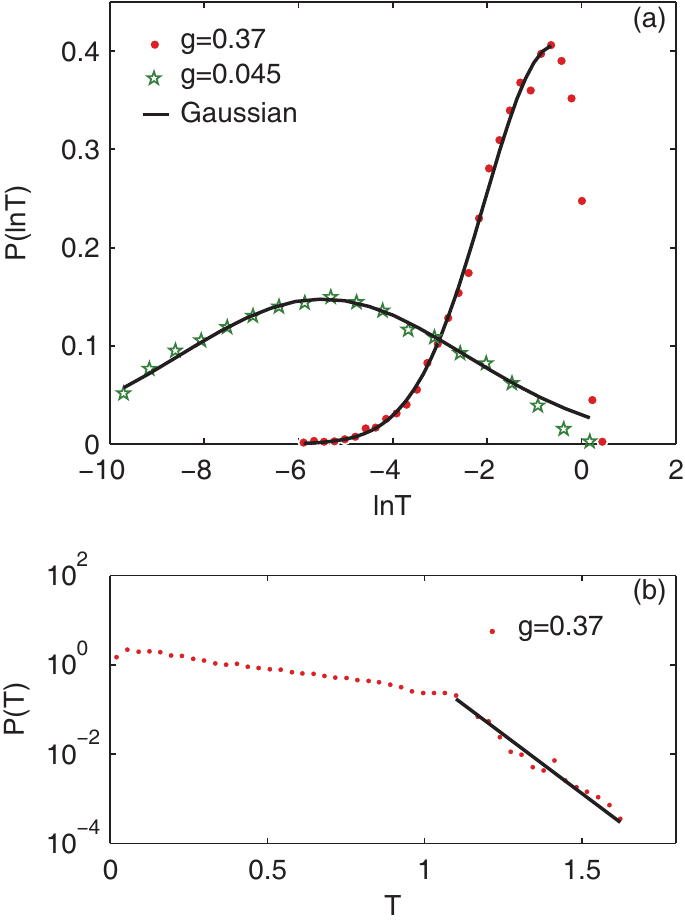}
\caption{Probability distribution of conductance. (a) $P(\ln T)$ for \textsl{g}= 0.37 (red dots) and 0.045 (green asterisk). The solid black line is a Gaussian fit to the data. For \textsl{g}= 0.045, all the data points are included, while for \textsl{g}= 0.37, only data to the left of the peak is used in the fit. (b) $P(T)$ for {\textsl g}= 0.37 in a semi-log plot exhibits an exponential tail.}\label{Fig1}
\end{figure}

These results can be understood with the aid of the Coulomb gas model \cite{32} for the eigenvalues of the TM based on Dyson's treatment of large random Hamiltonian matrices \cite{33}. In this model, the $\tau_n$ are associated with positions $x_n$ of parallel lines of charges and their images with the same sign of charge at $x_{-n}= -x_n$ embedded in a continuous charge distribution of opposite polarity. The transmission eigenvalues can be expressed in terms of the charge positions as, $\tau_n=1/\cosh^2(x_n)$. The repulsive interaction between charges, and hence between transmission eigenvalues, gives rise to universal conductance fluctuations for diffusive waves \cite{15,16,17,18}. For deeply localized waves, the large value of $x_1$ and its large separation from $x_2$ translates through the relation $\tau_n=1/\cosh^2(x_n)$ into transmission being dominated by the first transmission eigenchannel with $T\sim\tau_1\sim 4\exp(-2x_1)$. Since the distribution of $x_1$ for deeply localized waves is Gaussian \cite{32}, this leads to the log-normal distribution for $T$ for multichannel systems seen in Fig. 2a for $\textsl{g}=0.045$ and $L/\xi\sim4$. We show below that in addition to explaining the origin of the distributions of transmittance in the diffusive and localized limits, the charge model can explain the anomalous distribution of transmission in the intermediate case in the crossover between diffusive and localized transport. Understanding transport in terms of the charge model is further of importance because the charge model provides a parallel treatment of diffusive and localized waves and so forms the basis for a universal description of wave transport.

\begin{figure}[htcp]
\centering
\includegraphics[width=3.3in]{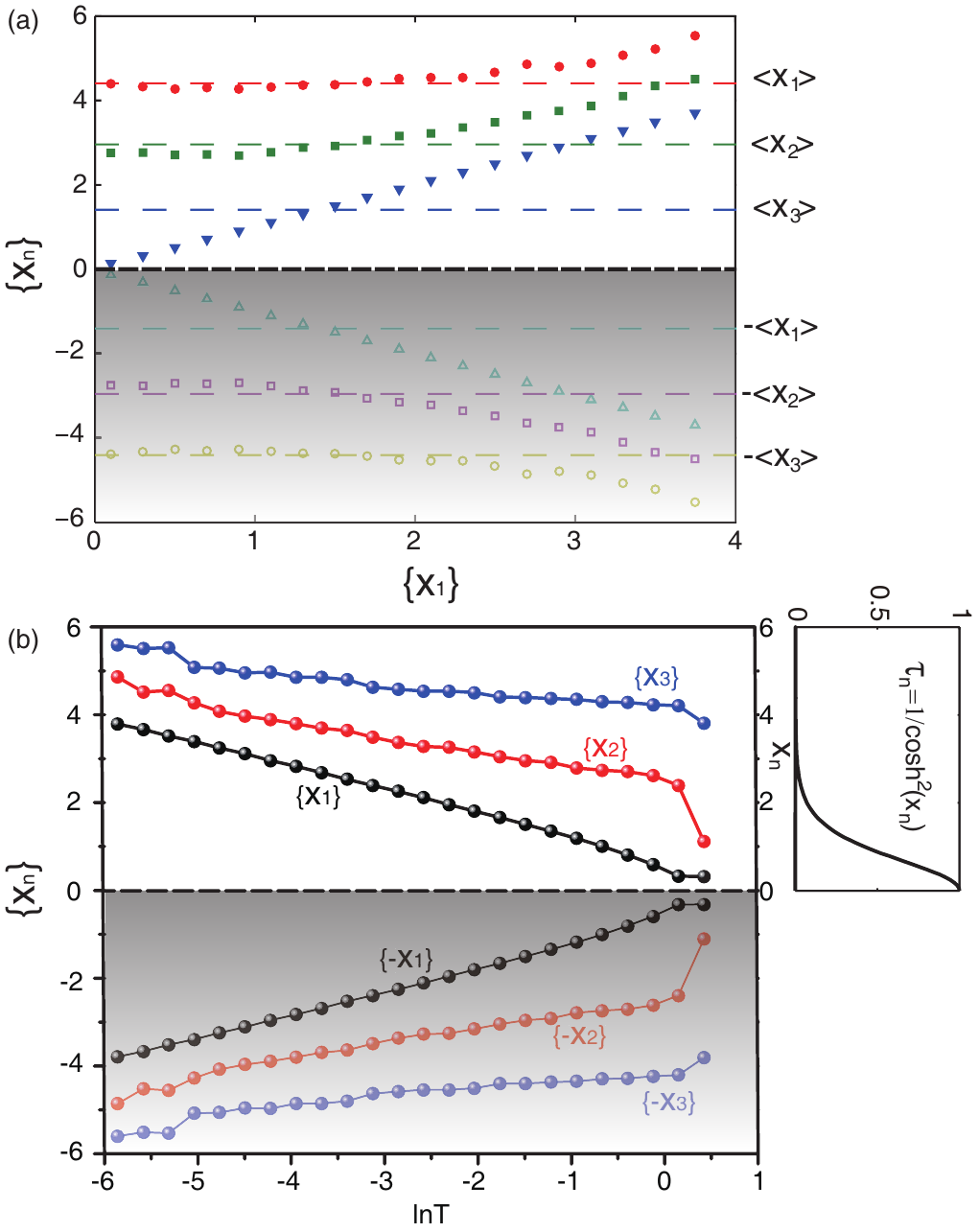}
\caption{Coulomb gas model of transmission eigenvalues and conductance. (a) Average positions of charges and their images with respect to different positions of the first charge $x_1$ in the random ensemble with {\textsl g}= 0.37. The dashed lines show the average positions of the charges for this ensemble. (b) Average positions of charges vs. $\ln T$ in the same ensemble. The curly brackets indicates the averaging is over subset of transmission matrices with the specified value of $\ln T$. The correspondence between $\tau_n$ and $x_n$ is presented in the side panel.}\label{Fig2}
\end{figure}
In Fig. 3a, we plot the variation of the average positions of the charges for different positions of the first charge $x_1$ in the random ensemble with \textsl{g}= 0.37. The repulsion between $x_1$, which is associated with the highest transmission eigenvalue $\tau_1$, and its image at -$x_1$ enforces a ceiling for $\tau_1$ of unity. The average spacing between $x_1$ and $x_2$ increases as the value of $x_1$ decreases since the charge at $x_2$ then interacts strongly with the charge at $x_1$ as well as with its nearby image at $-x_1$. At the same time, the spacings between $x_2$ and $x_3$ and their average positions hardly change. This reflects the tendency to heal large fluctuations in charge positions for more remote charges. The source of the sharp cutoff in $P(\ln T)$ can be seen in the context of the Coulomb gas model by examining the spacings of charges for different values of $\ln T$, as shown in Fig. 3b. A relatively high value of $T$ is only achieved when the first charge is near the origin. This is an unlikely circumstance because this charge is strongly repelled by its image. $P(\ln T)$ would be expected to fall off especially rapidly for values of $T$ above unity since this would requires two charges along with their images to be close together near the origin as seen in Fig. 3b. 

\begin{figure}[htbp]
\centering
\includegraphics[width=3.4in]{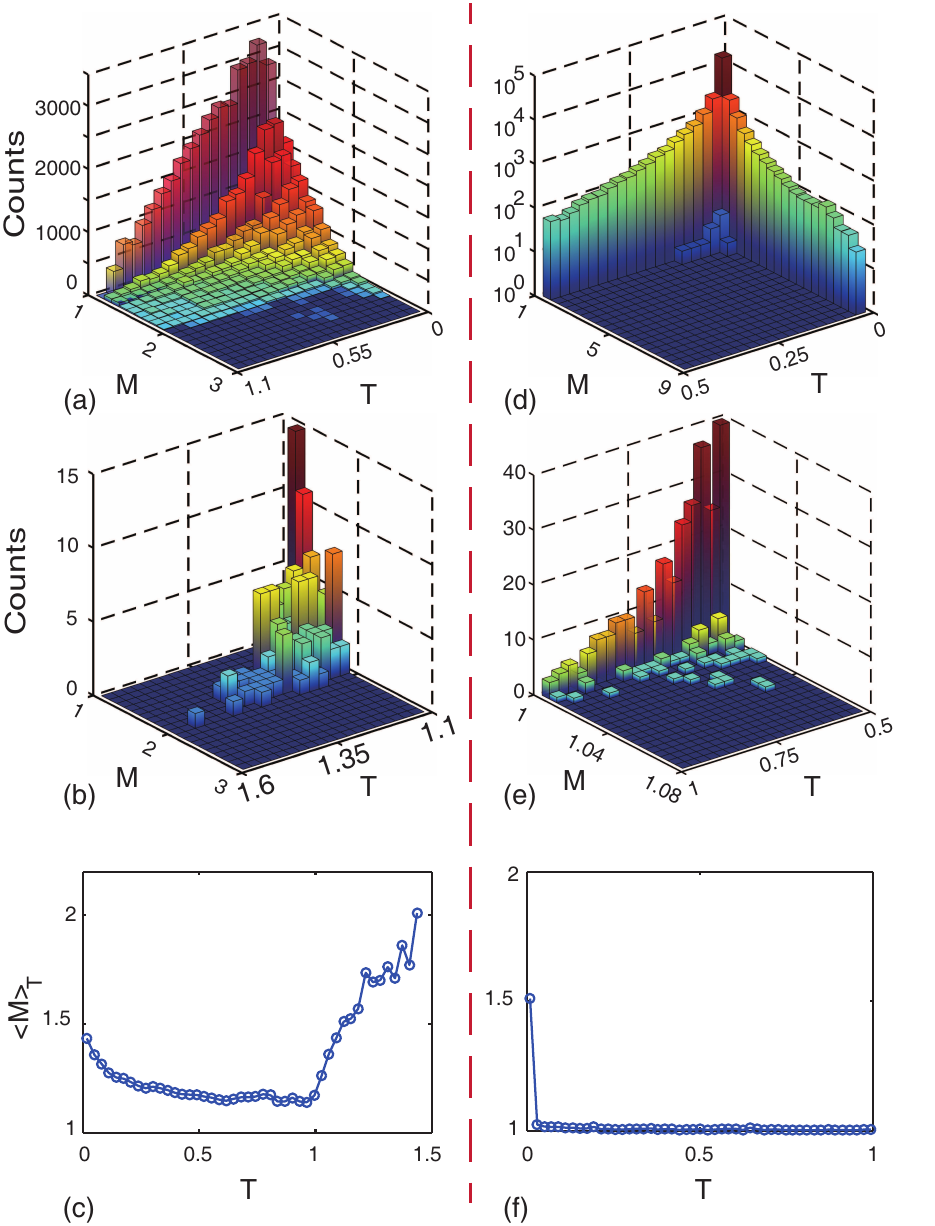}
\caption{The joint probability distribution of $T$ and $M$, $P(T,M)$, for two random ensembles of {\textsl g}= 0.37 (left column) and 0.045 (right column). (a) and (b) Joint distribution $P(T,M)$ for (a) $T< 1.1$ and (b) $T> 1.1$. (c) $\langle M\rangle_T$ for different $\langle T\rangle$ in the random ensemble. $\langle M\rangle$ increases to 2 for the highest value of transmittance. (d) and (e) Joint $P(T,M)$ for (d) $T<0.5$ and (e) $T>0.5$. (f) $\langle M\rangle_T$ vs. $T$ for the ensemble with ${\textsl g}=0.045$. High transmittance {\it T} is found when a single transmission eigenchannel contributes to transmission, $M\sim 1$.}\label{Fig3}
\end{figure}
We show below that the impact of the correlation between the set of $\tau_n$ and $T$ upon the probability distributions of conductance or conductance may be explained as well in terms of the joint distribution of $T$ and the participation number of transmission eigenchannels, $M \equiv (\sum_{n=1}^N \tau_n)^2/\sum_{n=1}^N \tau_n^2$. In addition, knowledge of the joint distribution of $P(T,M)$ is sufficient to fully describe the probability distributions of total transmission and intensity, $T_a$ and $T_{ba}$. Plots of $P(T,M)$ for {\textsl g}= 0.37 and 0.045 are presented in Fig. 4. A peak in $M$ occurs at low values of $T$ for localized waves since, typically, the incident frequency is then not on resonance with an electromagnetic mode of the sample. Several of the closest off-resonance modes, and hence several eigenchannels which are largely composed of these modes, contribute to transmission. For larger values of $T$, the frequency is often closer to resonance with a particular mode or with overlapping modes with similar speckle patterns. Transmission is then often dominated by a single channel with a speckle pattern similar to that of the resonant mode. Nonetheless, when $T>1$, more than a single channel must contribute to transmission as is seen in Figs. 4a-c. In contrast, in the more strongly localized samples, $\langle M\rangle$ is seen in Figs. 4d-f to fall sharply and remains close to unity for the highest values of $T$ observed. The highest values of $T$ observed in this ensemble are below unity.  

The ratio of var($\ln T$) and $-\langle \ln T\rangle$, ${R}$, vs. $L/\xi$ is plotted in Fig. 5 and seen to approach unity for $L\gg\xi\sim 24$ cm \cite{34}. This occurs just as $M$ approaches unity\cite{35}. The participation number of eigenchannels is represented in Fig. 4 as the average of $M$ weighted by {\it T}, ${\langle M T \rangle}/\langle T \rangle $ is shown in Fig. 5, as is the weighted average of $M^{-1}$, ${\langle M^{-1} T \rangle}/\langle T \rangle $. The latter average closely tracks $R$. Thus, the SPS is approached in a multichannel quasi-1D random system when transmission is through a single eigenchannel. In this limit, the statistics of transmission in a multi-channel quasi-1D waveguide, for which $L$ exceeds the width of a constant cross section with reflecting sides, is essentially the same as in 1D system. 
\begin{figure}[htc]
\centering
\includegraphics[width=3in]{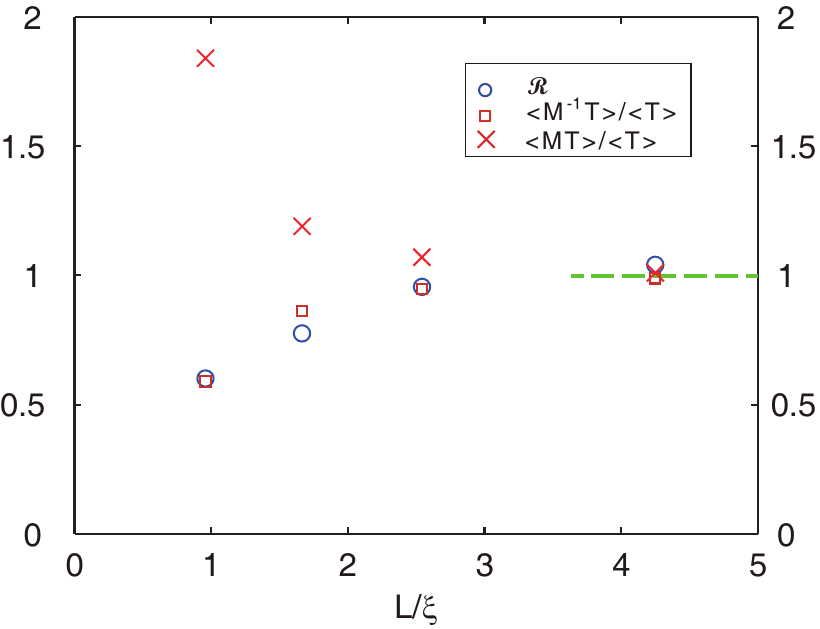}
\caption{Approach to single parameter scaling in multichannel random systems. ${R}\equiv-{\rm var}(\ln T)/\langle \ln T\rangle$, $\langle M T\rangle/\langle T\rangle$ and $\langle M^{-1} T\rangle/\langle T\rangle$ with respect to $L/\xi$. The dashed line is the prediction of SPS for large $L/\xi$.}\label{Fig4}
\end{figure}

\subsection{Statistics of intensity and total transmission}
In a single multichannel sample, $M^{-1}$ is equal to the variance of the total transmission for a single incident channel $T_a$ relative to its average in that configuration, $M^{-1}$=var$(NT_a/T)$ and the distribution of $(NT_a/T)$ has a form that depends only upon $M$ \cite{35}. This distribution changes from Gaussian to exponential as $M^{-1}$ varies from 0 to 1. The forms of these distributions are observed experimentally but an analytic expression has not been found as yet. Thus, the distribution of total transmission can be expressed as $P(T_a)=\int P(T,M)P(NT_a/T;M)dTdM$ and depends only on $P(T,M)$. Since, the distribution of intensity $T_{ba}$ normalized by the total transmission $T_a$, is a universal negative exponential \cite{36}, the statistics of intensity also depends only upon $P(T,M)$ and so on {\textsl g}.

\begin{figure}[htc]
\centering
\includegraphics[width=2.6in]{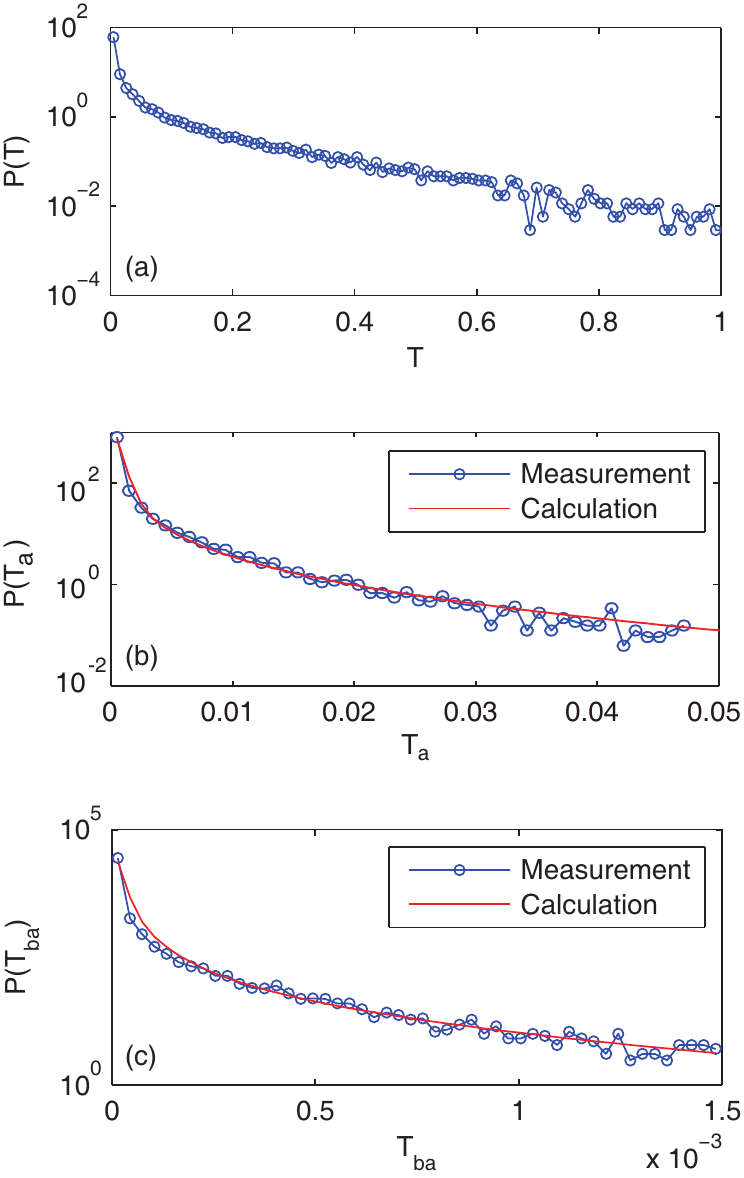}
\caption{Statistics of transmittance $T$, total transmission $T_a$ and intensity $T_{ba}$ for deeply localized waves with {\textsl g} = 0.045. (a) Probability distribution of {\it T}. (b) Distribution of $P(T_a)$ in the ensemble. Assuming $M=1$ for all configurations, $P(T_a)$ can be calculated as, $P(T_a)=\int P(NT_a/T)P(T)dT$ (red curve). (c) Intensity distribution $P(T_{ba})$ is calculated with the use of the calculated $P(T_a)$ in 6(b) and the universal negative exponential function of $P(NT_{ba}/T_a)$ and compared with measurements.}\label{Fig5}
\end{figure}
The relationship between the statistics of transmittance, total transmission and intensity is particularly straightforward for deeply localized waves for which $M\rightarrow1$, in nearly all transmission matrices except for those associated with the lowest values of {\it T}. This is seen in Figs. 4d-f and 5 to arise for {\textsl g}= 0.045. We can therefore obtain the total transmission and intensity distribution from the distribution of the transmittance in this case by setting $M=1$ and $P(T,M) =P(T)$ for values of $T>0.005$. The distribution of total transmission relative to its average value in a given sample realization at a specific frequency is then $P(NT_a/T)=\exp[(-NT_a/T)/(T/N)]$, and the distribution of relative intensity is $P(N^{2}T_{ba}/T)=\exp[(-NT_{ba}/T_a)/(T_a/N)]$. As a result, the distributions of intensity and total transmission in this limit depend only upon the distribution of $T$. This is demonstrated in Fig. 6 for the sample with ${\textsl g}=0.045$ and $\langle MT\rangle/\langle T\rangle=1.02$. Good agreement is seen between the distributions of intensity and total transmission and calculations in which $M$ is set to unity. 

Because the distribution of normalized intensity is a negative exponential, $P(NT_{ba}/T_a)=\exp(-NT_{ba}/T_a)/(T_a/N)$, the fluctuation of normalized intensity, $s_{ba}=T_{ba}/\langle T_{ba}\rangle$, is closely linked to the fluctuation of the normalized total transmission, $s_a=T_a/\langle T_a\rangle$,$\langle s_{ba}^n\rangle=n!\langle s_a^n\rangle$ \cite{37,38}. Similarly, in the limit of $M=1$, $\langle s_a^n\rangle=n!\langle s^n\rangle$, where $s=T/\langle T\rangle$. This gives the expression for the variance of $s_{ba}$ when transport is dominated by a single transmission eigenchannel, ${\rm var}(s_{ba})=2{\rm var}(s_a)+1=4{\rm var}(s)+3$. We found the values of the variances of $s_{ba}, s_a$ and {\it s} to be 27.5, 13.2 and 6.4 in the sample of $L$= 102 cm, which satisfies these relations. This further confirms that the transport is via a single eigenchannel in this random ensemble. Thus, SPS is approached within a random ensemble of random waveguides when transmission is through a single eigenchannel.

\section{Conclusions}
In conclusion, we have observed the transition to SPS and related this to the changing correlation between transmission eigenvalues and the transmittance and to the joint distribution of transmittance and the participation number of transmission eigenvalues. The correlation between the $\tau_n$ is interpreted in terms of a Coulomb gas model including image charges. A one-sided log-normal distribution of transmittance is observed at $L/\xi \sim$ 1 and a log-normal distribution for deeply localized waves at $L/\xi \sim$ 4. We find that the statistics of conductance and transmission approach SPS in quasi-1D samples as $\langle M\rangle \rightarrow 1$ in the limit $L/\xi\gg 1$. Since $P(NT_{a}/T)$ in a single TM depends only on $M$ \cite{35}, $P(T,M)$ determines the statistics of intensity, total transmission, and transmittance in any random ensemble. The role of $P(T,M)$ in local and integrated transmission is demonstrated for deeply localized waves. The results presented here demonstrate the power of the TM and the multichannel Landauer relation to unify the study of the statistics of electronic conductance and optical transmission.

\begin{materials}
\section{Experiment setup}
Measurements are carried out in collections of randomly positioned alumina spheres with a diameter of 0.95 cm and refractive index 3.14 embedded in Styrofoam shells. These spheres are contained within a copper tube of diameter 7.3 cm giving an alumina sphere volume fraction of 0.068. The transmission coefficient $t_{ba}$ between incident point {\it a} and output point {\it b} is measured with the use of wire antennas connected to a vector network analyzer. The source and detector antennas are mounted on a two dimensional translation stage so that they can move freely on the transverse dimension of the copper waveguide. The TM is obtained by measuring $t_{ba}$ between arrays of points on the incident and output surface for microwave radiation polarized along the orientation of the wire antenna. Measuring only a single polarization corresponds to a loss of control which modifies the probability distribution of transmission eigenvalues but does not substantially affect the transmittance distributions for localized waves (45). The dimensionality of the measured TM matches the number of channels of the copper tube, with {\it N} $\sim$30 in the frequency range 10-10.24 GHz. The sample tube is rotated and vibrated after each measurement of the full TM to produce a new realization of the random sample. Measurements of spectra of the TM are made for sample lengths {\it L} = 23, 40, 61 and 102 cm in 23, 60, 45 and 50 sample configurations, respectively. 

\section{Recursive Green's function simulation}
Since the distribution of normalized transmittance $s=T/\langle T\rangle$ depends upon only the value of the dimensionless conductance {\textsl g} of the ensemble of random waveguides and not upon details of the structures, we determine the value of {\textsl g} in our samples by comparing the distribution of normalized transmittance obtained from Green's function simulations of transport of scalar wave in random quasi-1D samples with semi-infinite ideal leads with a dielectric constant of $\epsilon_1$= 2.25 and perfectly reflecting transverse sides to our measurements. The disordered region is modeled by a position-dependent dielectric constant $\epsilon(x,y)=\epsilon_2+\delta \epsilon (x,y)$. The wave equation $\triangledown^2E(x,y)+k_0^2\epsilon(x,y)E(x,y)=0$ is discretized using a 2D tight-binding model on a square grid and solved with use of the recursive Green's function method (46). Here, $k_0$ is the wave number in the sample leads. In the simulations, $\epsilon_2$ is equal to 2.25  and $\delta \epsilon$ is chosen from a uniform distribution in the range [-1.1,1.1]. The wavelength of the incident wave in the simulation is 2$\pi$/$\sqrt{\epsilon_1}$ measured in units of the lattice spacing. The width of the waveguide is 60 and the length varies from 300 to 1450 in units of the lattice spacing. The number of propagating waveguide modes {\it N} is equal to 32. By comparing $P(\ln s)$ between simulations and measurements, we estimate the values of {\textsl g} is 0.65, 0.37, 0.19 and 0.045 for the sample length {\it L}= 23, 40, 61 and 102 cm. For weakly localized samples, the value of {\textsl g} found is close to 2/3var($s_a$), in which $s_a$ is the normalized total transmission, $s_a=T_a/\langle T_a\rangle$.

\section{Removing the impact of absorption on the statistics of transmission}
The impact of absorption on the statistics of transmission is removed by Fourier transforming the field spectrum into the time domain and multiplying the time signal by exp({\it t}/2$\tau_a$), where {\it t} is the time delay and 1/$\tau_a$ is the absorption rate (7). The validity of this procedure is confirmed in recursive Green's function simulation with and without an imaginary part of the random dielectric function. After compensating for the loss associated with the imaginary part of the dielectric function in a random ensemble with {\textsl g}= 6.7, we find that the corrected distribution of transmittance is Gaussian with the same mean as the sample without absorption and a variance which is 3\% larger. 
\end{materials}





\begin{acknowledgments}
We thank Matthieu Davy for stimulating discussions regarding the statistics of the transmission matrix. We also thank Arthur Goetschy and A. Douglas Stone for the simulation program and Howard Rose for help with the experimental setup. The research was supported by the NSF under Grant No. DMR-1207446.
\end{acknowledgments}





\end{article}









\end{document}